\documentclass[%
 reprint,
 amsmath,amssymb,
 aps,
]{revtex4-1}

\usepackage{graphicx}
\usepackage{dcolumn}
\usepackage{bm}

\begin{document}

\preprint{APS/123-QED}

\title{Detection of self-generated nanowaves on the interface of an evaporating sessile water droplet}

\author{Dhanush Bhatt}
\affiliation{ 
Indian Institute of Technology, Madras, India, 600036
}%
\author{Rahul Vaippully}%
\affiliation{ 
Indian Institute of Technology, Madras, India, 600036
}%
\author{Bhavesh Kharbanda}
\affiliation{ 
Indian Institute of Technology, Madras, India, 600036
}%
\author{Anand Dev Ranjan}
\affiliation{ 
Indian Institute of Technology, Madras, India, 600036
}%
\author{Sulochana R.}
\affiliation{%
Indian Institute of Science Education and Research, Trivandrum, India
}%
\author{Viraj Dharod}
\affiliation{%
Indian institute of Technology, Roorkee, India
}%
\author{Basudev Roy}
 \email{basudev@iitm.ac.in}
\affiliation{ 
Indian Institute of Technology, Madras, India, 600036
}%

\date{\today}

\begin{abstract}
Evaporating sessile droplets have been known to exhibit oscillations on the air-liquid interface. These are generally over millimeter scales. Using a novel approach, we are able to measure surface height changes of 500 nm amplitude using optical trapping of a set of microscopic particles at the interface, particularly when the vertical thickness of the droplet reduces to less than 50 $\mu$m. We find that at the later stages of the droplet evaporation, particularly when the convection currents become large, the top air-water interface starts to spontaneously oscillate vertically as a function of time in consistency with predictions. We also detect travelling wave trains moving in the azimuthal direction of the drop surface which are consistent with hydrothermal waves at a  different combination of Reynolds, Prandtl and Evaporation than previously observed. This is the first time that wave-trains have been observed in water, being extremely challenging to detect both interferometrically and with infra-red cameras. We also find that such waves apply a force parallel to the interface along the propagation direction.
\end{abstract}

\pacs{Valid PACS appear here}
\maketitle


\section{Introduction}
As the size of electronic and optoelectronic devices shrink into the micro and nano-scales, and the power density rises, problems of heat and mass dissipation into the surroundings attain more and more prominence. There can be heat transport in many ways, namely radiative which have been shown to be very high when dissipation is directly into a gas with comparable mean free path \cite{radiative}, conductive \cite{conductive} or convective \cite{convective}. Convective heat transport is still one of the most powerful ways of dissipating energy particularly when the surface area becomes large compared to the volume of the object. The exact mechanisms of such dissipation at the microscale are only beginning to emerge. One such facet is the formation of hydrothermal waves at the sub-micron scales during heat transport. 

Hydrothermal waves (HTWs) \cite{craster} are thermally induced travelling waves occurring on a gas-fluid interface and are particularly relevant to the cases of heat and mass transfer in the microscopic domain like in cooling of miniature chips or medical diagnostics on biofluid samples \cite{biofluid1}. These obtain their energy from temperature gradients which can either be externally applied onto fluid layers \cite{davis,smith, davis1} or even spontaneously generated due to evaporation of the fluid at the interface\cite{craster}. These HTWs propagate almost perpendicularly to the temperature gradient at large Prandtl numbers, Pr = $\mu C_p/\lambda$, where $\mu$ is the kinematic viscosity, $C_p$ is the specific heat at constant pressure of the fluid and $\lambda$ is the thermal conductivity.

There have been reports of detection of HTWs on evaporating sessile droplets of volatile liquids like ethanol and methanol, which have Pr of about 4 and cease to be travelling for Fluorocarbons which have lower Pr \cite{craster} which have subsequently been explained partially \cite{sefiane} as well. It was reported in \cite{sefiane} that hydrothermal wavetrains could be explained using mode stability analysis at Reynolds number of about 100 (Re = $\frac{\rho U H_0}{\mu}$, where, U = $\epsilon \gamma \Delta T/\mu$, $\epsilon$ = $H_0/R_0$, $\gamma$ is the surface tension gradient at ambient temperature, $\rho$ is the density of the fluid, $H_0$ is the height of the sessile liquid droplet, $R_0$ is the transverse radius of the droplet and $\Delta$T is the temperature difference between the liquid and gas) with a Pr of 1 and Evaporation number (Er = $\frac{\lambda \Delta T}{\mu L Re}$, where L is the Latent heat of evaporation of the fluid) of 0.0005 which ceased to be travelling and became evaporative rolls at Re of 20, Pr of 7 and Er of 0.0005. Detection of HTW waves in water, having higher Pr of about 7, has proved elusive, given that the temperature difference at the top and the substrate is less than 1 C \cite{craster}, not to mention that the Infra-Red (IR) cameras \cite{Niliot} used only have a spatial resolution of 8 $\mu$m in the far field. It is here that we use a new novel strategy relying upon optical trapping \cite{ashkin} of microscopic particles at the air-water interface and then detecting vertical displacements to show the existence of HTWs at Re of 32, Pr of 7 and Er of about 0.006, the conditions in which only evaporative rolls have been reported earlier \cite{sefiane} and also where the cut-off travelling wave mode number is less than 10. We show that HTWs of about 500 nm amplitude propagating in the azimuthal direction to the circular sessile droplet are detected using this technique having a mode number much higher than 10, given as $\frac{2 \pi  500}{34}$ (about 100), where the transverse extent of the droplet is about 500 $\mu$m and the wavelength of the travelling wave is 34 $\mu$m. We also show for the first time that these waves can apply forces on particles placed at the interface, quite akin to a ball placed on a sea-wave, which can subsequently be used for mass transfer at the nanoscale. 

As mentioned in \cite{sefiane}, a detailed model that would quantitatively match with these experimental observations requires full three-dimensional dynamic simulations, since the flows are inherently three-dimensional. Further effects like vapor concentration in the gaseous phase and substrate thermal properties would have to incorporated into the model to make better explanations for the observations. Such modelling is beyond the scope of the present manuscript.

\section{Experimental details}

We use a thorlabs optical tweezers kit OTKB/M for the tweezers apparatus. It focuses the light to a diffraction limited spot via a Nikon E-Plan 100x/1.25 Numerical Aperture (NA) oil-immersion objective, where the particle is trapped, and the forward scattered light is collected via a Nikon E-Plan 10x/0.25 NA condenser. The first trapping laser used here is a 976 nm butterfly laser from Thorlabs Inc. The second laser of Lasever, 1064 nm wavelength diode laser, is used when the second trap is present. The light collected is made incident on a 1064 nm-pass interference filter (1064 $\pm$ 5 nm) to separate two different wavelengths of 976 nm and 1064 nm, and then the 1064 nm light is made incident an edge mirror to collect the two halves of the beam on separate photodiodes. The photodiode currents were amplified using current preamplifiers, PDA200C (Thorlabs). A visible white light LED illuminates the sample from the top via a dichroic mirror and is detected at the bottom behind another dichroic mirror using a CMOS camera (thorlabs).
 The X motion is inferred from the difference in signals from the two photodiodes while the Z motion is inferred from total intensity of the two photodiode signals.

\begin{figure}
\includegraphics[width=\linewidth]{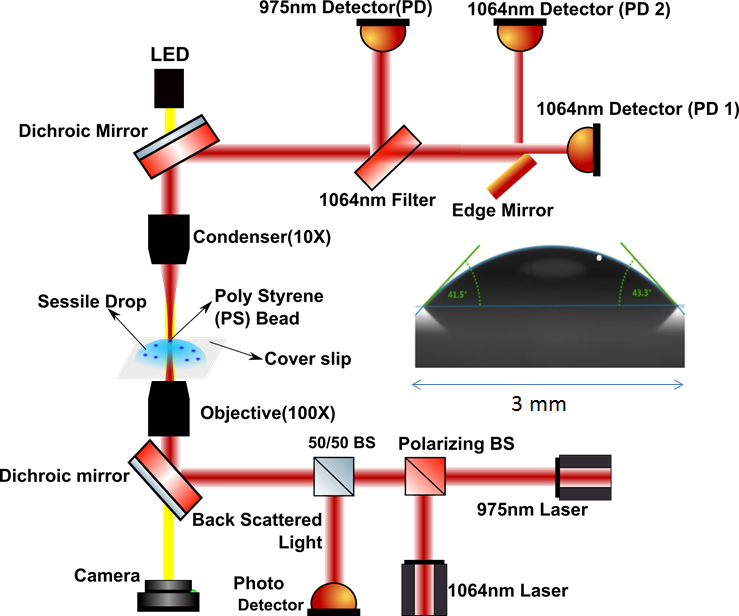}
\caption{\label{schematic} This figure describes the schematic of the experiment. Two lasers, one at 1064 nm and the other at 976 nm have been combined at a polarizing beam splitter and focused into a diffraction limited spot using a 100x oil-immersion objective to generate two spots in proximity to each other on an air water intererface of a sessile evaporating water droplet. The scattered light is collected in the forward direction using a 10x condenser and then split into the two wavelength components using an interference filter 1064 nm. An edge mirror is then placed in the path of the 1064 nm light so as to split the light to detect x motion and detected into two photodiodes as the difference in signals between PD1 and PD2. The z intensity is inferred from the total intensity over the two photodiodes PD1 and PD2. The 976 nm total intensity provides information about the z displacement of the particle trapped in that light. The sample is illuminated using a white light LED and mixed at a dichroic mirror, and also extracted from the trapping light at the bottom using another dichroic mirror. The inset shows the sessile water droplet freshly dropped onto a glass cover slip for the experiment. The glass sample is hydrophilic such that the contact angle is less than 90 degrees. The backscatter signal is split using a 50/50 beam splitter in the backward direction. Only one laser is used when the backscatter is detected. The double wavelength laser detection in backscatter could not be performed here due to low levels of scattered light and cross talk.}
\end{figure}

The water droplet (de-ionized water) of a fixed volume (typically 3 $\mu$l), with a dilute suspension of polystyrene particles of  3.1 $\pm$ 0.2 $\mu$m diameter (Thermo Fisher Scientific), is dropped onto a glass coverslip of typical thickness 120 $\mu$m (no. 1 size, English glass) , mounted on an XYZ stage with micrometer screws to position them. The coverslip is hydrophilic in nature with a contact angle of about 43 degrees for a water droplet placed on it (shown in the inset of Fig. \ref{schematic}.  The stage also has a provision of a piezo-electric transducer for electronic control in all the three dimensions. The coverslip is mounted above the 100x objective and the water forms a free boundary on the top with air. The temperature of the room is maintained with an air-conditioner at 27 C. The experiment was repeated many times over a span of 3 months without active humidity control. The relative humidity varied from 60 to 70 percent. Care was taken such that there were no external air currents around the sample. The air-conditioner was placed at the far side of the room, away from the apparatus. 

Trapping of particles in such open configurations is difficult because the working distance of the 100x, 1.25 NA objective is 170 $\mu$m while the water droplet on glass can extend by more than that. Therefore we allow the water to gradually evaporate so as to have the interface to within 50 $\mu$m such that particles can be trapped on them. Trapping on air-water interface has been performed in a Langmuir trough using lower NA objectives (NA of 1) which have a larger working distance \cite{interface1}, but require higher power to trap. We show that even higher NA objective can also be used. The polystyrene particle is trapped just under the interface using the tweezers light and the activity of the interface studied. 

The 976 nm and the 1064 nm trapping laser beams are coupled into the sample chamber after passing it through a polarizing beam splitter. The laser intensities are adjusted to have about 30 mW in the sample plane each. These lasers are infra-red lasers which do not interact with water, not to mention that the power is low enough not to heat it by more than 0.01 C. Further, the laser beam has been focused into the top interface of the water which that it does not interact too much with the bulk of the medium. Thereafter, the light just escapes the water medium. Thus, in view of these aspects, namely the low intensity and the wavelength of the laser, the effect on convections currents due to the light is minimal. The reflected light emerges via the second port of another 50:50 beam splitter placed in the path, thereby enabling detection of the back-scattered light. Thus we have provisions to detect both the forward and backscattered signals. .

 We trap the polystyrene particle just under the interface. We find that as the interface continues to dry, and the convection currents on the surface start to increase upon formation of temperature difference between the substrate and the top of the interface due to evaporation, the vertical interface also starts to oscillate. 

\section{Results and Discussions}

\begin{figure}
\includegraphics[width=\linewidth]{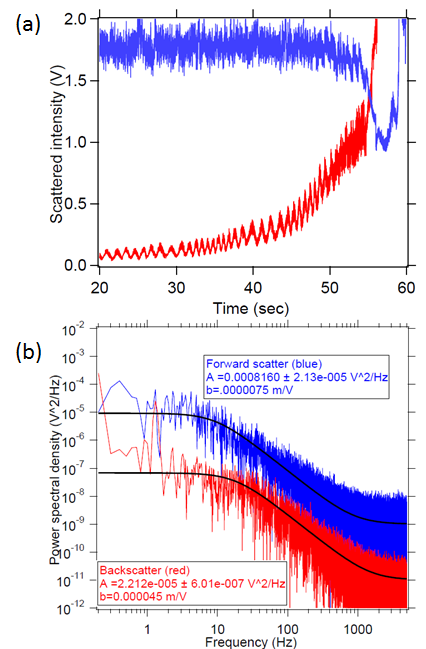}
\caption{\label{scatter}(a)This figure indicates the forward-scattered (blue) and back-scattered (red) intensities for the same 3.1 $\mu$m diameter silica particle trapped at the air-water interface of an evaporating water droplet. As the water evaporates and the vertical thickness of the droplet becomes smaller and smaller, the convection currents become stronger and stronger. It is at this stage that the vertical interface of the drying droplet also starts oscillating. The forward scatter is less sensitive to these oscillations while the backscatter is more sensitive. At about 55 seconds on this time series, the particle escapes from the trap upon the influence of the strong currents. Some surface oscillations are visible in the forward scatter too, particularly when the amplitude becomes large (b) The corresponding power spectral densities that indicate the calibration factor for the measurements. The forward scatter measurement seems to have more noise than the corresponding backscatter measurement for the same particle. The surface oscillations at about 1 Hz are noticable in the PSD. The trapping laser is the 976 nm one with about 20 mW at the sample plane. }
\end{figure}

We can only trap the particle when the water has dried enough to allow the vertical interface to be within about 50 $\mu$m. At this point, the water starts to evaporate very rapidly such that we typically get about 2 minutes to do our experiments after which the convection currents become too strong to allow trapping. We show a set of typical forward and backward scattered signals off the trapped particle in Fig. \ref{scatter} , just before the convection currents become too strong. The blue curve in Fig. \ref{scatter}(a) indicates the forward scattered signal while the red curve indicates the backscatter signal. At about 55 seconds on the time series, the convections currents become very strong and drag the particle out of the trap. However, it still allows probing the system till that time. We find that the backscatter signal indicating the vertical motion of the particle starts to oscillate thereby showing local vertical displacements. This is expected as the evaporation of water causes the interface of the sessile droplet to oscillate. However, the forward scattered light detects very little of this motion. We suspect that this happens due to the nature of the signals in the forward and backward directions \cite{pralle,avin}, the forward scatter being an interference between unscattered light and the scattered light, while the backscattered one being interference between that reflected from the interface and that scattered off the particle. In the case of low amplitude oscillations, the interface oscillates while the particle remains quite unaffected, such that the backscattered detection senses it while the forward scatter does not. However, we succeed in detecting oscillations in forward scatter when the amplitude of the oscillations becomes very large, particularly late in the evaporation process because the motion of the interface induces motion of the particle itself in the vertical direction. We show the power spectra of these times series in Fig. \ref{scatter}(b), where the extra noise in forward scatter signal is apparent, possibly due to more light collected. The power spectrum of the backscatter signal shows a peak the oscillation frequency as well. The power spectra have been fitted to a lorentzian of the form of eq. \ref{psd} \cite{erik}.

\begin{equation}
    y =\frac{A}{1+(f/f_c)^2}
    \label{psd}
\end{equation} 

where, A is the amplitude of the lorentzian, f the frequency and f$_c$ is the corner frequency. From here the calibration factor for converting the Volts into position displacement is obtained using the relation in eq. \ref{calib}.

\begin{equation}
    b = \sqrt{\frac{k_B T}{A \gamma}}
    \label{calib}
\end{equation}

where, k$_B$ is the Boltzmann constant, T the temperature, and $\gamma$ is the translational drag coefficient. The drag coefficient for motion along the direction perpendicular to a perfectly slipping interface, like in the case of an air-water interface is given by eq. \ref{faxen} \cite{lauga}.

\begin{equation}
    \gamma = \frac{\gamma_0}{1-\frac{3a}{4h}}
    \label{faxen}
\end{equation}

where, $\gamma_0$ is the drag coefficient away from any interfaces, a is the radius of the particle and h is height from the center of the particle to the interface. Thus very close to the interface, this drag coefficent shall be $4\gamma_0$.

\begin{figure}
\includegraphics[width=\linewidth]{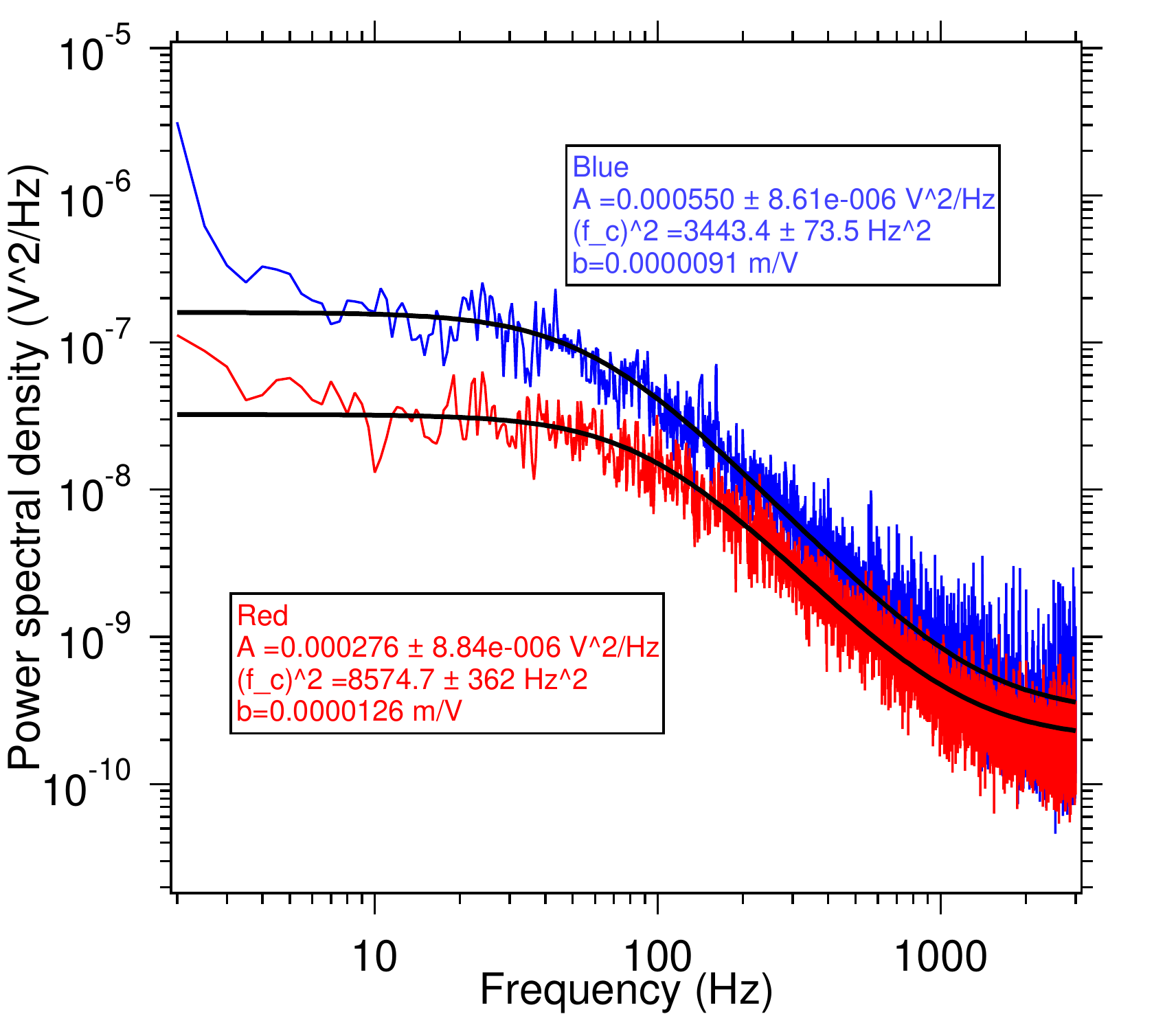}
\caption{\label{fft_two} This figure indicates the power spectral density for the vertical motion of the two particles used to measure waves on the interface of air and water. The particles are trapped very close but under the interface and displaced in the transverse direction by about 18 $\mu$m, such that the viscosity is still that of water. The red curve corresponds to that of 976 nm laser while the blue, 1064 nm. Each beam has about 50 mW laser power. The corresponding calibration factors for the z-signal have also been estimated here, shown as b. The A is the amplitude of the lorentzian, given in eq. \ref{psd}, while the f$_c$ is the corner frequency. The trap stiffness is about 0.017 pN/nm for the blue curve (the 1064 nm laser), while the red curve (the 976 nm laser) has 0.026 pN/nm. }
\end{figure}

\begin{figure}
\includegraphics[width=\linewidth]{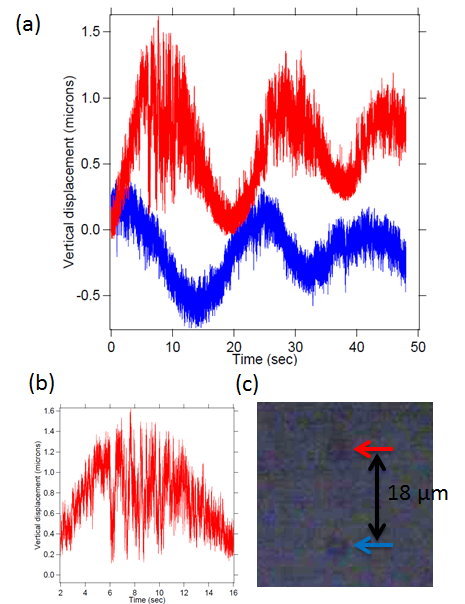}
\caption{\label{wave1}(a) The vertical displacements of the two particles trapped in 976 nm (red) and 1064 nm (blue) lasers. The particles are placed azimuthally with a separation of 18 $\mu$m between them. The time traces have been displaced in the ordinate axis for a better view. (b) The red trace zoomed-in to highlight bistability between the vertical interface and the trap. (c) A snapshot of the two 3.1 $\mu$m diameter polystyrene particles trapped on the air-water interface of the evaporating sessile water droplet, 18 $\mu$m away tranversely from each other along the y direction.}
\end{figure}

We move to another question, whether these disturbances indicate standing waves or travelling waves. In order to answer this question, we add an extra trap in the vicinity of the first trap using another laser. Then we place two particles next to each other and see how the vertical disturbances at the two spots look like. This has been indicated in Fig. \ref{wave1}. The corresponding power spectra are indicated in Fig. \ref{fft_two}, which have also been used for calibration of the amplitude of disturbances. The trap stiffness for the 1064 nm laser is 0.017 pN/nm while that of 976 nm laser is 0.026 pN/nm. 

\begin{figure}
\includegraphics[width=\linewidth]{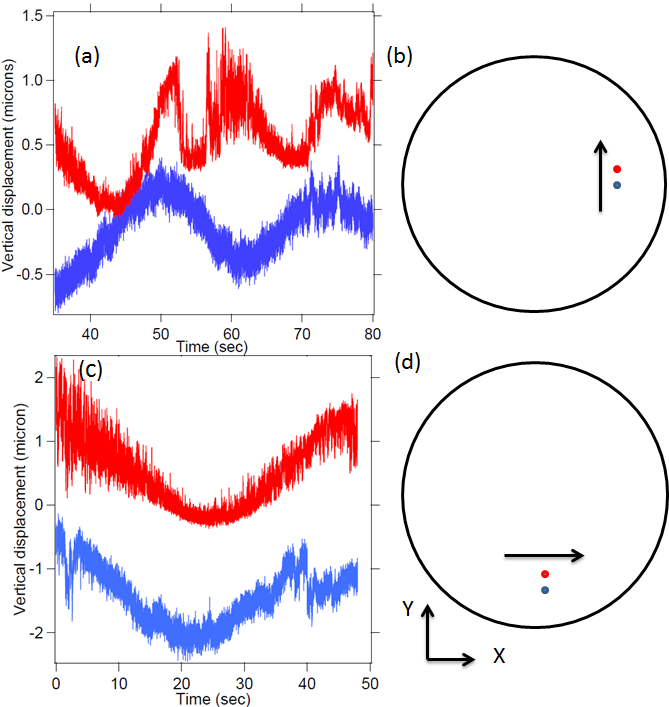}
\caption{\label{wave2} The vertical oscillation of two particles placed transverse to each other along (a) azimuthal direction (c) radial direction. The low phase delay along the radial direction indicates that the waves are propagating along the azimuthal direction. The cartoons indicate the positioning of the particles on the water droplet as viewed from above, (b) is azimuthal and (d) is radial, to ascertain the wave direction. The wave is propagating in a slightly different direction than the azimuth and thus the phase delay is not completely 0. However, estimation of the propagation direction would be difficult from here as each of these measurements have been performed on different sets of particles
which may not have the same propagation speed along radial and azimuthal directions. The red time trace corresponds to a particle that has been placed along the red dot, while the blue trace corresponds to the blue dot. The arrow shows the approximate direction of the wave-train. The red and the blue time traces in each curve have been displaced along the ordinate axis for clarity. The diameter of the water droplet on glass is about 3 mm. }
\end{figure}

The fig. \ref{wave1} indicates disturbances at two locations on the interface of the droplet which form a line parallel to the edge of the interface (the azimuthal configuration, indicated in Fig. \ref{wave2}(b)). We find that there is a phase lag between the two disturbances which is neither 0 degrees nor 180 degrees. This indicates travelling waves on the surface of water. Such waves have been noticed in sessile evaporating droplets of ethanol, methanol and fluorocarbons \cite{craster}, explained as hydrothermal waves and detected using IR cameras. However, so far, nobody has successfully demonstrated this in water using infra-red cameras, possibly due to lower temperature variations at the crests and troughs. We detect these waves using our two-trap system and even ascertain that the amplitude of these disturbances on water is about 0.5 $\pm$ 0.2 $\mu$m. The height fluctuations look very similar to that theoretically predicted in Fig. 7(a) of \cite{sefiane}. Given that the delay between the waves is about 10 seconds, and the separation between the particles is about 18 $\mu$m between the centers, we calculate that the speed of this wave is about 1.8 $\mu$m/sec. Then the wavelength of the wave is about 34 $\mu$m, given that the period of the wave here (Fig. \ref{wave2}(b))) is 19 seconds. It may be noted here that \cite{sefiane}
mentions mode numbers less than 10 being visible with the infra-red cameras, in the conditions they have analysed. We show that much higher mode numbers are getting detected here. We need a detailed quantitative three-dimensional simulation to understand this. 

These waves have also been dfficult to detect using IR cameras because the minimum resolution of the camera in the far-field configuration is about 8 microns, such that the crest would only appear as 1 to 2 pixel wide, hardly enough to be called a proper signal. Further, these waves are also hard to detect interferometrically, to do which one shall have to reflect part of the light from the water-glass interface at the bottom and part from the top air-water interface which  would be about 50 $\mu$m apart in the best of cases. Finding objective lenses that can have a Rayleigh range of 50 $\mu$m in conjugation to having less than 10 $\mu$m spot size is a very challenging configuration. Thus, our two-trap technique provides a good way to address the problem, where we can detect disturbances higher than 200 nm over large ranges of about 30-40 $\mu$m.   

We also note in the red curve of Fig. \ref{wave1}, that there is higher noise particularly at higher displacements. We suspect that this is due to the vertical interface going above the position of the tweezers. Even at the vertical interface, the particle feels a low potential region, such that the particle now has a vertical double well potential effectively. Then it exhibits bistability in this potential. It is this that we highlight in our Fig. \ref{wave1}(b). We see that the noise suddenly increases towards the top of the curve when the trap gets sufficiently displaced from the interface.  

We now address the question of the direction of the travelling wave. We place the two-particle system in two configurations, namely, azimuthal, Fig. \ref{wave2}(a), highlighted in (b), and radial, Fig. \ref{wave2}(c), highlighted in (d). The we study the delay betweent the two sets of signals in either configuration. We find that the delay is very small in the radial direction while it is substantial in the azimuthal direction. The separation between the 3.1 $\mu$m diameter polystyrene particles used here is about 18 $\mu$m. This seems to indicate that the wave is propagating almost along the azimuthal direction. There can small deviations from this direction as indicated by the radial phase delay. However we cannot estimate the precise value because these experiments were performed during different iterations with different droplets, such that we are not sure whether the speed of the wave and the period shall be the same. We do find that the periods are quite variable from one iteration to the next. 

\begin{figure}
\includegraphics[width=\linewidth]{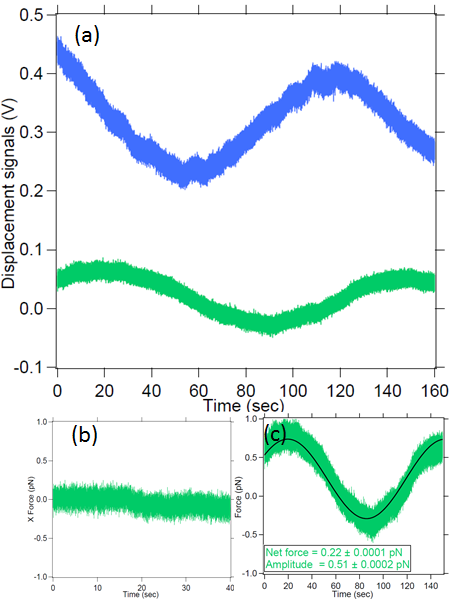}
\caption{\label{force} This figure indicates the generation of traverse force by the wave. (a) The blue curve indicates vertical interface oscillations while the green curve indicates the corresponding transverse x-motion of the same particle trapped at the blue spot on the air-water interface for the configuration in Fig. \ref{wave2}(d). (b) A reference curve showing the x displacements when the interface does not oscillate (c) The x-direction force, corresponding to a particle in the case of (a), in the presence of vertical oscillations. The curve has been fit to a sinusoid. This shows that the wave not only carries energy but also applies a transverse force on a particle placed on the interface to the extent of 0.22 pN in this case. }
\end{figure}

We go on to test whether such waves can actually apply force in the direction parallel to the interface, quite akin to sea-waves. This can then have potential applications in mass transfer. For this, we track the x displacement simultaneously with the vertical (z) displacement and find that indeed there is a phase delay between the z and the x signals, as indicated in Fig. \ref{force}(a). The blue curve in this figure refers to the z-displacement while the green curve indicates x-displacement. We also calibrate the force applied by the wave and show the results in Fig. \ref{force}(c), while the x-displacement in absence of waves has been indicated in Fig. \ref{force}(b). The calibrated x force in the presence of the wave shows a shift from 0 by about 0.22 pN, obtained as the offset of the sinusoidal curve fit to the x-signal, which might be the force that a wave can apply on the particle towards mass transfer.  

These nanowaves were obtained in water at a Re of 32, Pr of 7 and Er of 0.006, quite in contrast with previous measurements on ethanol and methanol \cite{craster}, which were done at Re of about 100, Pr of 4 and Er of 0.0005. The main difference between water and ethanol or methanol during spontaneous evaporation is the extent of the temperature differential created at the same values of room and substrate temperature. In the case of water, it can be at most 1 C while methanol or ethanol can acquire about 4 to 5 C. This changes the evaporation number Er. The wavelength of the waves observed in methanol was more than 1 mm while that in water in the present manuscript is about 34 $\mu$m. As has been mentioned earlier, the exact quantitative match with simulations require detailed three-dimensional modelling which is beyond the scope of the present manuscript. 

\section{Conclusions}

Thus to conclude, we detect disturbances on the air-water interface of a sessile evaporating droplet at a room temperature of 27 C using a particle trapped with optical tweezers. We find that the backscatter detection is more sensitive to these surface deformations than the forward scattered light. We explore further that these disturbances are due to spontaneous waves developed on the interface upon influence of evaporation and subsequent hydrothermal wave formation. These waves are hard to detect either with thermal imaging via infra-red cameras or even with interferometric techniques. We find that these wave propagate in the azimuthal direction. We also find that this wave displaces the particle in the direction parallel to the interface and applies a force of about 0.22 pN on a particle. Such a technique can be used to study the dynamics of interfaces of drying liquid droplets at the nanometric level with high speeds, thereby providing a wealth of information about the type of material and the processes involved. Such technique can also be useful for study of nanowaves on elastic surfaces like cells and polymers under different conditions.

\begin{acknowledgments}
We wish to acknowledge the support of IIT Madras for their seed grant and Dillip Satapathy for helpful comments on the manuscript. 
\end{acknowledgments}

\nocite{*}
\bibliography{main}

\end{document}